\documentclass[runningheads]{llncs}

\usepackage{graphicx}

\newcommand{\etal}[0]{\textit{et al.}}

\begin{document}

\title{Control in Hybrid Chatbots}

\titlerunning{Control in Hybrid Chatbots}

\author{Thomas Rüdel\inst{1} \and Jochen L.~Leidner\inst{2,3,4}\orcidID{0000-0002-1219-4696}}

\authorrunning{T.~Rüdel \and J.~L.~Leidner}

\institute{Kauz GmbH, Erasmusstraße 15, 40223 Düsseldorf, Germany\and
  Information Access Group, Coburg University of 
  Applied Sciences and Arts, Friedrich-Streib-Straße 2, 96450 Coburg, DE \and
  University of Sheffield, Regents Court, 211 Portobello, Sheffield S1 4DP, UK \and
  KnowledgeSpaces UG (haftungsbeschr.), Erfurter Straße 25a, 96450 Coburg, DE\\
  Contact: \email{thomas.ruedel@kauz.net,leidner@acm.org}}

\maketitle

% -----------------------------------------------------

%\begin{abstract}
%  Customer data typically is held in database
%  systems, which can be seen as rule-based knowledge base,
%  whereas businesses increasingly want to benefit from the
%  capabilities of large, pre-trained language models. 
%
%  In this industry presentation, we describe the case study
%  of how a commercial rule engine and an integrated neural
%  chatbot are integrated, and what level of \emph{control}
%  that particular integration mode leads to.
%  We also discuss alternative ways (including past ways
%  realized in other systems) how researchers strive
%  to maintain control and avoid what has recently been called
%  model ``hallucination''.
%  
%  \keywords{search \and recommendation \and conversational search \and chat-bots \and conversational agents \and natural language understanding and generation \and explainability and accountability \and control \and ``hallucination'' \and hybrid rule-based and
%  statistical/neural systems \and retrieval applications.}
%\end{abstract}

% -----------------------------------------------------

\vspace{-.5cm}

\section{Introduction}

\textbf{Motivation.}
Chatbots and AI-agents have become widespread in customer service and in applications like knowledge management, recommender systems, and help desks. Businesses increasingly want to benefit from the capabilities of large language models like OpenAI’s GPT-4 and applications powered by such models.
 Nevertheless, the use of generative AI by companies has been seriously slowed down by concerns about
data protection and by the fact that generative AI is known to sometimes make things up -- create ``hallucinations'' as it is often called. Even if an answer does not contain
hallucinated information, it may still suffer from incompleteness or misleadingly
connected pieces of information. However, companies that want to use AI-agents in non-trivial circumstances need to be able to control them, in particular in customer-facing applications. It would be very unfortunate if it misinforms customers about the company’s
products or prices. It should also stick very closely to the intended marketing messages.

While there is a lot of discussion about ``safe AI'', ``reliable AI'', ``trustworthy AI'', ``explainable AI'' (XAI) etc., the question of ``controllable AI'' is rarely discussed. However, as stated above, it is very often crucial that enterprises cannot just rely on, 
but are in fact able to control 
an AI system (more precisely, exercise control at design time how the system will behave at runtime).

\textbf{Background.}
We define \emph{controllable} AI as an AI system the behavior of which a systems engineer can
control to a certain degree. In this sense, control is not an absolute concept -– a totally controllable AI system would be equivalent to a deterministic system -- but rather a relative one: the
engineer may be able to exert higher or lower levels of control over an AI system. From this definition it should also be clear that control does not mean that the \textit{user} is able to
control the AI system but rather that the \textit{designer} is able to control how the AI system
behaves towards the user.

The starting point for all approaches is the use of \emph{Retrieval-Augmented Generation} (RAG) as a paradigm. RAG -- in combination with appropriate prompting -- tries to ensure that an AI system answers users’ question solely on the basis of data provided by the company, and \emph{not} on the basis of its training data, which may include a lot of irrelevant or even misleading internet data. Within this paradigm both the retrieval and the generation step can be designed to allow for more or less control over the output. In section 4 we will see several architectural options
how a range of control-retaining architectures can be designed.

So is controllable AI the same as reliable AI? No, as we have seen, reliable means delegation to an AI system hoping for mostly positive outcomes. Controllable means actively directing and steering AI systems, i.e. where the agency lies is shifted to the system designer \cite{Pizzi-Scarpi-Pantano:2021:JBusRes}. RAG is one form of
controllable AI, and often a useful first step.
Pizzi et al. \cite{Pizzi-Scarpi-Pantano:2021:JBusRes} raises the important
question of ``Who has the control when interacting with a
chatbot?'', and the answer should be: the system engineer should
design the system so that they are in control, and in a way so
that the user feels in control, too.
The highest level of control is provided by a purely
rule-based system. The lowest of control is provided by using an LLM "as is", such as by typing a question directly into the ChatGPT user interface. Neural language
models typically have a built-in temperature
parameter that controls their degree of reliance on
randomness.\footnote{Setting it to 0 should force
these models to emit strictly the most-likely
next word. In practice, it has been observed that
there often remains residual randomness \cite{Chann:2023:online}, which is undesirable from
a business perspective.}
Medium levels of control combine sophisticated search on both text and meta-data, ranking of results and careful prompt design for generation. In effect, "generation" effectively becomes summarizing the most relevant documents. 

Clearly, however, more control requires more effort/cost, so there will always be a tradeoff between the two.
% Role of the Client

In this paper and associated industry presentation, we describe the case study of how a commercial rule engine and an integrated neural language model are combined, and what level of \emph{control}
that particular integration mode leads to.
We also discuss alternative ways (including past ways realized in other systems) how researchers strive to maintain control and avoid what has recently been called ``hallucination'' \cite{Jha-etal:2023:ICAA}.

\section{Related Work}

The term \emph{chatbot} is a business term rather than a scientific
one, and is used to refer to a ranger of technically rather distinct software
artifacts \cite{Adamopoulou-Moussiades:2020,Luo-etal:2022:WIDM} (Folstad, Skjuve and Brandtzaeg \cite{Folstad-Skjuve-Brandtzaeg:2019} present a typology of chatbots). In this section, we briefly review the relevant technologies.

\subsection{Natural Language Interfaces to Structured Databases}

Androutsopoulos, Ritchie and Thanisch \cite{Androutsopoulos-Ritchie-Thanisch:1995:NLE} give a readable survey of answering questions
against databases up to 1995, so before the recent ML paradigm shift.
Leidner and Kamkova \cite{Leidner-Kamkova:2013:FQAS} showed that
the task of question answering over structured data can be solved by
generating an index
of question-answer pairs at indexing time that then get found by
a trivial ranked retrieval instead of following the canonical
pipeline (question type classification, question analysis,
question to SQL query translation etc.). They demonstrate the efficacy
on an example use case of searching for toxic spills.
Plachouras \etal{} \cite{Plachouras-etal:2016:SIGIR} applied their
method for question answering over macroeconomic indicators (e.g.
\textit{What was the GDP of India in 2012?} and financial deals
(M\&A).
Li and Jagadish \cite{Li-Jagadish:2015:VLDB} present
a recent syntax-directed method for translating questions into SQL.

\subsection{TREC-Style Question Answering Systems}

U.S. NIST conducted a series of annual shared tasks on open-domain
question answering (1998-2007) where question answering was
framed as answer extraction of different types of questions
(factoid questions, definition questions, list questions) from text \cite{Hickl-etal:2006:TREC,Leidner-etal:2003:TREC}.

\subsection{Spoken/Written Dialogue Systems}

Spoken dialog systems interact with users in the spoken medium \cite{Jokinen-McTear:2022,McTear:2002:CSUR}, but,
like (good) chatbots, they need to maintain state
and context, which they accomplish using a dialog
manager \cite{Lee-etal:2010:JCompSciEng}.
Henceforth, we are only concerned with chatbots that target
providing information relevant to the customers of a business;
another kind of chatbot that aims to behave indistinguishable from
humans (Turing test) is not within our scope.

\subsection{Modern Hybrid Systems, Hallucination \& Control}

Adiwardana \etal{} \cite{Adiwardana-etal:2020:ArXiv}
introduce the chatbot \emph{Meena} (by Google Brain)
and the evaluation metric SSA, the average of
``sensibleness'' and ``specificity'' as judged by
humans. Meena was shown to perform favorably when
compared against previous chatbots like
%Cleverbot\footnote{\url{}},
DialoGPT \cite{Zhang-etal:2020:ArXiv},
%Mitsuku\footnote{\url{}} or
XiaoIce \cite{Zhou-etal:2020:CL}.
Both Weston \etal{} \cite{Weston-etal:2018:EMNLP} and
Lewis \etal{} \cite{Lewis-etal:2020:NeurIPS} introduce a general-purpose fine-tuning recipe for RAG, an approach to overcome the limited knowledge of foundational pre-trained language models by integrating search results.
Lewis and co-workers report that for generation tasks, RAG models generate more specific, diverse and factual language than
earlier parametric-only sequence-to-sequence
approaches.
Mao \etal{} \cite{Mao-etal:2021:ArXiv}
proposed the inverse of RAG,  Generation-Augmented Retrieval (GAR) and applied it to  answering open-domain questions.
Maeng and Lee \cite{Maeng-Lee:2021:AsCHISympos}
explore three alternative architectures for their
chatbot targeting victims of crimes, namely a rule-based,
learning-based and hybrid architectures, respectively.
Luo \etal{} \cite{Luo-etal:2022:WIDM} discuss
various alternative hybrid architectures for chatbots.

Pande, Martin and Pimmer \cite{Pande-Martin-Pimmer:2023:AAAISprSympos} discuss dialog management
strategies that include pre-trained language models in the context of a medical coaching system.
Notably, they involve a mechanism for human escalation, which may be seen as yet another useful
form to maintain control.
Jha \etal{} \cite{Jha-etal:2023:ICAA} and Rawte, Sheth and Das \cite{Rawte-Sheth-Das:2023:ArXiv}
provide surveys about language model hallucination.
Leiser \etal{} \cite{Leiser-etal:2023:MuC} conduct a participatory
design study to elicit requirements that language models should satisfy,
many of which were confirmed difficult to implement by machine learning
experts.

To our topic, work by Weston \etal{} \cite{Weston-etal:2018:EMNLP}, Lewis \etal{} \cite{Lewis-etal:2020:NeurIPS}
and Roller \etal{} \cite{Roller-etal:2021:EACL}
is most relevant: the former two works
introduced (between them, three forms of)
Retrieval Augmented Generation, on which we also rely,
whereas the latter compares three types of chatbot architectures, which the authors call ``retrieve'' (polyencoder architecture), ``generative`` (standard seq2seq transformer), and ``retrieve-and-refine'' \cite{Weston-etal:2018:EMNLP}.

% tutorial: Cai \etal{} \cite{Cai-etal-2022:SIGIR}

\section{Platform Description: A Basis for Hybrid Chatbots}

\textbf{Overall Architecture.} The overall architecture of the Kauz platform 
consists of components for design and analysis of AI-assistants, including a content management system and a design tool for dialogues, a runtime that manages conversations, supported by an LLM Manager and Kauz’ own NLU-Engine, interfaces to company data and backend-systems, and frontends for customer- and employee assistants (Fig. 1). Only the content management and design system and the runtime are relevant for explaining the concept of ``controllable AI'', so the other components will not be described further.
\begin{figure}
    \centering
    \includegraphics[width=\textwidth]{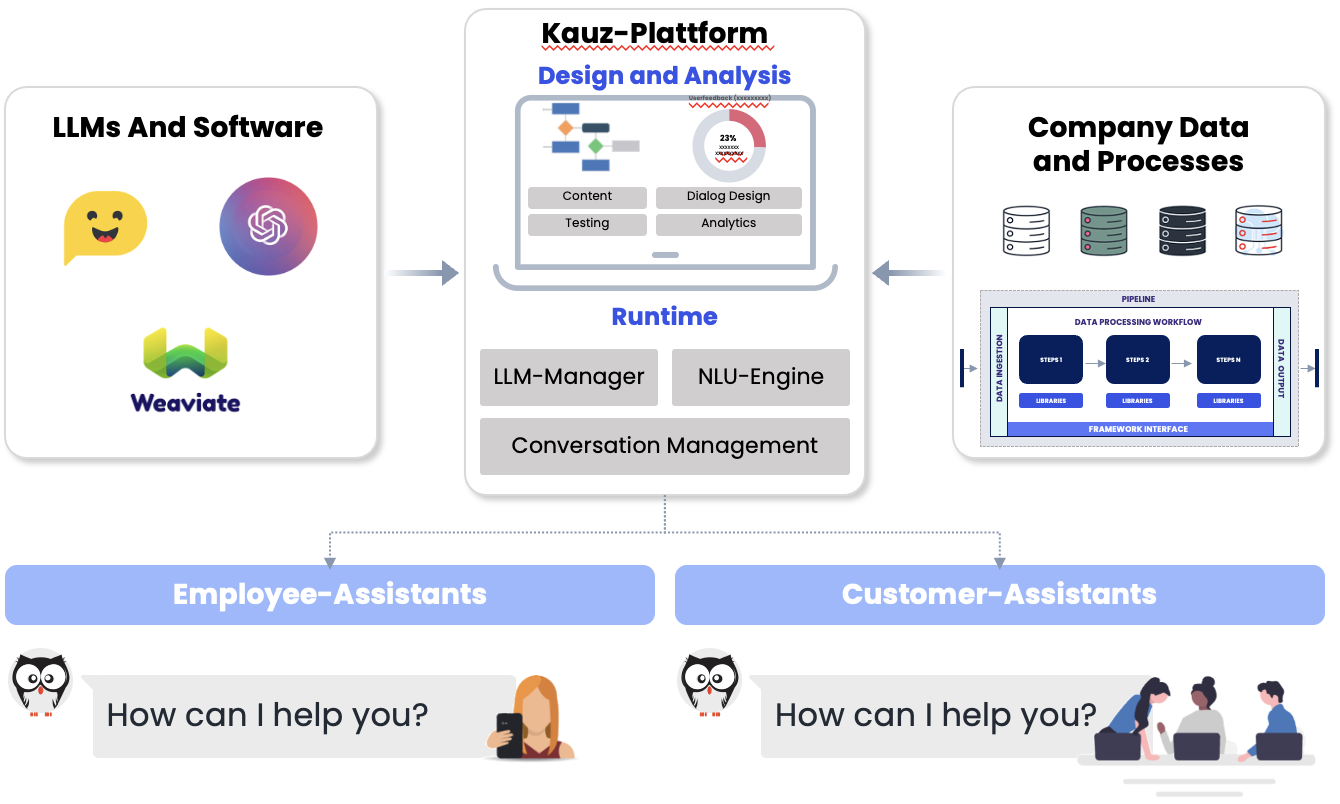}
    \caption{Architecture of the Kauz Platform}
    \label{fig:control2}
\end{figure}
The content management and design system enables (human) platform editors to load, edit, and delete documents from various sources, internal and external, but also manually add, edit, and annotate knowledge for AI-assistants. By design, it restricts AI-assistants to answer questions and execute tasks solely based on the data specifically provided by the company, giving it a very high level of control over the knowledge base of the assistant. In addition, it puts several tools into the hand of the (human) content editor that enable him to steer the assistant towards desired content and style of interaction with users. These tools, in order of increasing degree of control they enable are: prompt design, Annotation of documents with metadata, configurations for weighing metadata and text in selecting context for generation, using the Kauz NLU Engine to find the answer and suppress the generation mechanism altogether. The runtime consists of an LLM manager, the Kauz NLU engine, and a conversation management that integrates these components. The LLM manager can be configured to work with different LLMs, including GPT4 but also various open source models. It also integrates a vector database (Weaviate) and Langchain document loaders. \\

\textbf{Rule-Based Subsystem.} The Kauz NLU engine is a rule-based system with many of the expected elements, and a few unique features that distinguish it from more traditional systems. At its core is a NLP and search pipeline consisting of a parser -- the NLP component comprising a pre-processor, morphological analyzer, chunker, and clause integrator -- and an analyzer that evaluates terms, looks for referents for pronouns and referential NPs, searches information and formulates answers. This pipeline is supported by an ontology, lexicons, a grammar, and \emph{fact sheets}. The latter are a unique feature of the Kauz NLU engine: together with the ontology and some content from lexicons, they provide a knowledge graph of entities and states-of-affairs that AI assistants can draw on. Factsheets can represent individuals (like people, organizations or events), but also processes (like account openings and address changes), collections of such entities, and kinds. This enables a distinction for example between address changes in general (ontologically a kind), and the user’s address change in particular (an individual). The ontology in combination with the different lexicons create the possibility of a broad built-in language understanding, including synonyms, paraphrases, generalizations, and implications. A distinguishing feature of the system is that the analysis keeps track of how well a question and a given piece of information fit. If for example the question is about a product with a specific property – say chocolate containing nuts – and information is only found about chocolates but not about their content, it will be able to reflect the fact that is has found “supportive” but not “conclusive” information in its answers. This matching is somewhat similar to the confidence level that is used in machine-learning based models, but it is not probabilistic. The system ``knows what it does not
know'' would be an appropriate way to describe this situation. This fact allows a choice in designing the interplay between the NLU Engine and the LLM Manager, which can either be called in all circumstances where the NLU Engine has not found conclusive information, or only when the NLU Engine has not found relevant information at all. This choice depends a lot on the specifics of the use case and the relative detail of the available information bases.\\
\textbf{GenAI Substem (LLM Manager).}
The GenAI subsystem works with the paradigm of Retrieval Augmented Generation (RAG), so the input is not posed directly to the LLM, but is augmented with results of an intermittent search process. This ensures that the result is more likely to be on-topic. We argue that the use of the language model as a language model (in charge of verbalizing the answer, not sourcing the answer) is superior to the approach of sourcing answers from the LLM, which sometimes works and sometimes does not, depending on the nature of the training data.

The search component uses ranked retrieval based on the vector space model. It can be configured to analyze only the main text of documents or to include metadata and annotation. The latter possibility increases the level of control over the search outcome and therefore the final answer. The top-$k$ answers are appended to the query when passed on to the LLM Manager. Finally an LLM is provided with the augmented query as its input, and it is tasked with articulating a human language answer or recommendations, which are returned to the front-end.

\section{Control in Some Architectural Variants}

In this section, we describe some architectural variants in order to discuss finding a balance
between responses that impress with eloquence (LLM) and such responses
that stay factual, on-message, and avoid hallucination, such as those
retrieved from customers documents. As stated in the introduction, the starting point for controllable AI must be the Retrieval Augmented Generation (RAG) paradigm. Within this paradigm, both the retrieval and the generation steps can be designed to allow for more or less control over the output (Fig. 2). 
\begin{figure}
    \centering
    \includegraphics[width=.8\textwidth]{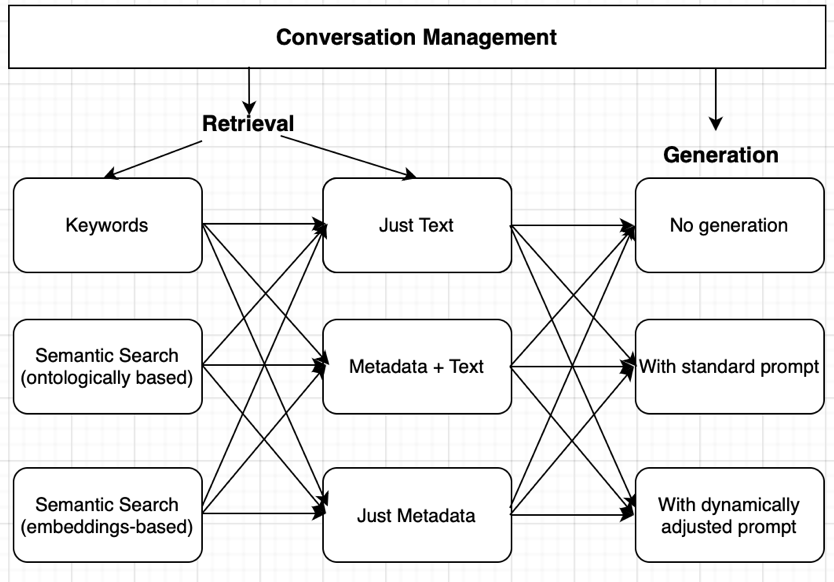}
    \caption{Conversation Management: Combining Methods}
    \label{fig:control}
\end{figure}

On the retrieval side, the key levers are the use of metadata or annotation, the weights given to these as opposed to just the text in the document, and of course the method used. The latter ranges from full-text search over ontology-based semantic search to vector-based search by an LLM. 

On the generation side, the spectrum ranges from unrestricted generation, triggered just by a standard prompt, via dynamic prompting based on the kind of input and context, to no generation for very sensitive information. 

Combining these levers in different ways yields a spectrum of operational models (Fig. \ref{fig:control2}). For example, using only metadata for retrieval and applying no generation yields a classic rule-based Q\&A system with maximum control. Using just vectorized text for retrieval and a standard prompt with an LLM for generation yields a comparatively low level of control. Combinations in between allow to adjust and fine-tune the approach to optimize the  combination of control achieved and effort required (since higher levels of control in general need more effort to set up the system appropriately). For example, a company may decide to prepare and annotate answers and dialogues for the most important questions and leave answering the long-tail rest of questions to the LLM, requiring only that the answer is based on the company’s knowledge base. 

The Kauz platform was designed with the objective to give AI-engineers and therefore the companies they work for a broad range of options about the degree of control they can exert over the AI’s behavior by combining the different approaches to retrieval and generation.

\begin{figure}
    \centering
    \includegraphics[width=\textwidth]{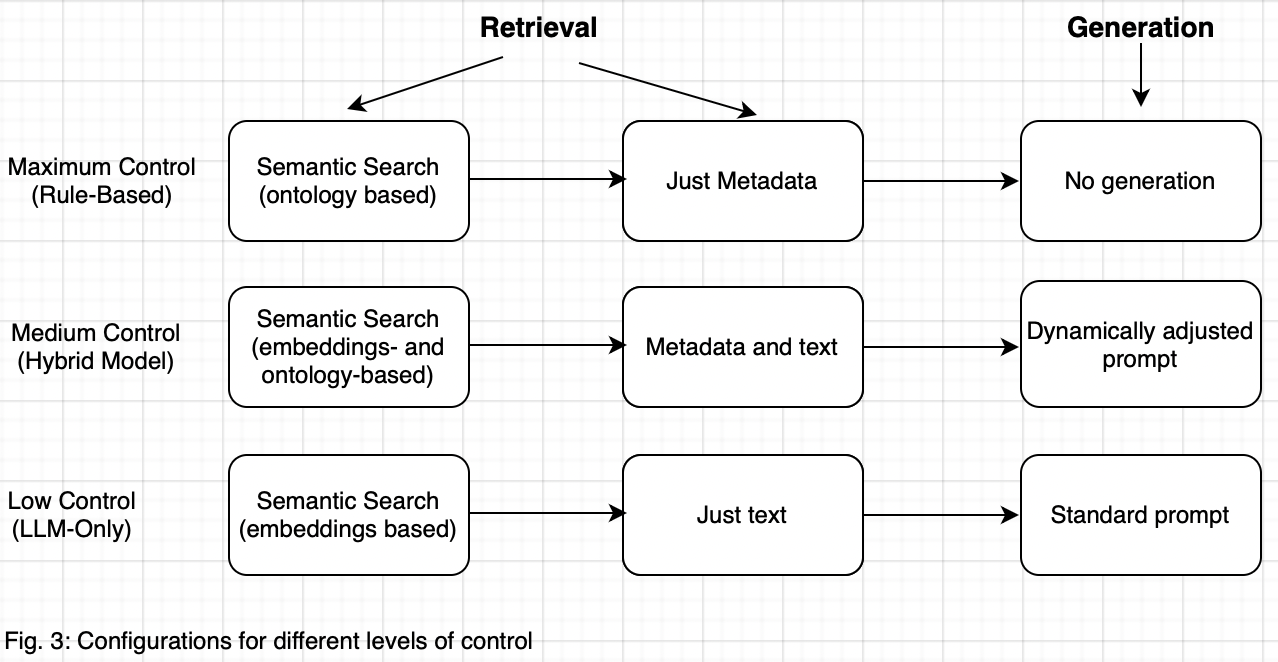}
    \caption{Architectural Variants for Retaining Control}
    \label{fig:control2}
\end{figure}

\section{Overall Lessons Learned}

In its work with customers, the Kauz team has acquired some
valuable lessons, some of which we would like to share:
\begin{itemize}
\item ``Retrieval is more important than generation'' (Table \ref{tab:mantra}).
\item The customer is \emph{not} the user, but the AI-engineer or chatbot editor.  It is the customer that needs
      to be happy, more so than the user (ideally, this coincides, but not
      always).
% \item \todox{ TBD - one more? }
\item To support the customer (i.e. the AI-engineer or chatbot editor), it is worth building out internal analytics functionality. We have
      built special B2B client analytics solutions to let our business
      clients that buy the platform rate the answers. Feedback from end-users is far less helpful, they use this option rarely, and mostly when frustrated.
\end{itemize}

\begin{table}[]
    \centering
    %\caption{Executive Summary}
    %\label{tab:mantra}
    \begin{tabular}{|lll|}\hline
         & \textbf{Take-Home Message} &\\ \hline
         & For many applications, retrieval is more important than generation - as long as all & \\
         & answer material is 
          sourced from company documents, hallucination and factual &\\
         & errors are unlikely or of minor consequence - the typical worst case is a lack of &\\
         & relevancy. &\\ \hline
    \end{tabular}
\end{table}

\section{Summary and Conclusion}

In this talk, we discussed the problem of maintaining control on the range
of answers that a chatbot designed for a particular domain, e.g. 
e-commerce, may provide to its user. We discuss a range of methods, 
building on the RAG paradigm, that can help steer the output towards the in-domain ``answer
space'' away from out-of-domain responses or ``hallucinations'' \cite{Jha-etal:2023:ICAA}, and we proposed a new architecture for selective LLM invocation that we hope will provide a balance between eloquently
formulated responses and absence of hallucinations.

% -----------------------------------------------------

\clearpage

\hrule

\vspace{.2cm}

\begin{appendix}

\section{Company and Authors}

\subsection{About the Company}

\textbf{About the Company.}
Kauz GmbH is a B2B software company founded 2016 in Düsseldorf, Germany, which integrates generative AI and state-of-the-art search and rule-based NLU to help companies build controllabe AI-assistants for customers and employees. 
\subsection{About the Speakers.}
\textbf{Thomas Rüdel.} Dr. Thomas Rüdel is the founder and CEO of Kauz GmbH. He studied economics and obtained a Ph.D. in econometrics. Prior to starting Kauz, he worked for twenty years as a  management consultant, and senior partner at McKinsey \& Company, advising clients across several industries. 

\subsubsection{Jochen L. Leidner}

Prof. Jochen L. Leidner, M.A. M.Phil. Ph.D. FRGS currently is the Research Professor for Explainable and Responsible Artificial Intelligence in Insurance at Coburg University of Applied Sciences, a Visiting Professor in the University of Sheffield. He is also founder and CEO of the consultancy KnowledgeSpaces.
His experience includes positions as Director of Research at Thomson Reuters and Refinitiv (A London Stock Exchange Group member) in London, where he headed its R\&D team, which he founded He was also the Royal Academy of Engineering Visiting’ Professor of Data Analytics at the University of Sheffield. His background includes a Master's in computational linguistics, English and computer science (University of Erlangen-Nuremberg), a Master's in Computer Speech, Text and Internet Technology (University of Cambridge) and a PhD in Informatics (University of Edinburgh), which won the first ACM SIGIR Doctoral Consortium Award.
\vspace{1cm}

\hrule

\end{appendix}

% -----------------------------------------------------

\clearpage

\bibliographystyle{splncs04}
\bibliography{hybrid-chatbot}

% \nocite{*}

\begin{figure}
    \centering
    \includegraphics[width=7cm]{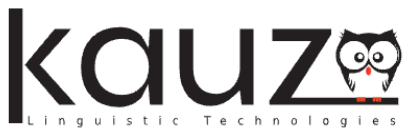}
\end{figure}

\end{document}